\documentclass[fleqn,11pt]{wlscirep}
\usepackage[utf8]{inputenc}
\usepackage[T1]{fontenc}
\usepackage{bm}
 \usepackage{setspace}
 \usepackage{lineno}
 \usepackage{chemformula}

\title{Homochiral antiferromagnetic merons, antimerons and bimerons realized in synthetic antiferromagnets}

\author[1]{Mona Bhukta}
\author[1,2,\textbf{*}]{Takaaki Dohi}
\author[1]{Venkata Krishna Bharadwaj}
\author[1]{Ricardo Zarzuela}
\author[1,3]{Maria-Andromachi Syskaki}
\author[4]{Michael Foerster}
\author[4]{Miguel Angel Niño}
\author[1]{Jairo Sinova}
\author[1,\textbf{*}]{Robert Frömter}
\author[1,\textbf{*}]{Mathias Kläui}

\affil[1]{Institute of Physics, Johannes Gutenberg-University Mainz, 55099 Mainz, Germany}
\affil[2]{Laboratory for Nanoelectronics and Spintronics, Research Institute of Electrical Communication, Tohoku University, 2-1-1 Katahira, Aoba, Sendai 980-8577, Japan}
\affil[3]{Singulus Technologies AG, Hanauer Landstrasse 107, 63796 Kahl am Main, Germany}
\affil[4]{ALBA Synchrotron Light Facility, 08290 Cerdanyola del Vallès, Barcelona, Spain}
\affil[*]{ \textcolor{blue}{tdohi@tohoku.ac.jp, froemter@uni-mainz.de, klaeui@uni-mainz.de}}

\begin{abstract}
The ever-growing demand for device miniaturization and energy efficiency in data storage and computing technology has prompted a shift towards antiferromagnetic (AFM) topological spin textures as information carriers, owing to their negligible stray fields, leading to possible high device density and potentially ultrafast dynamics. We realize, in this work, such chiral in-plane (IP) topological antiferromagnetic spin textures, namely merons, antimerons, and bimerons in synthetic antiferromagnets by concurrently engineering the effective perpendicular magnetic anisotropy, the interlayer exchange coupling, and the magnetic compensation ratio. We demonstrate by three-dimensional vector imaging of the Néel order parameter, the topology of those spin textures and reveal globally a well-defined chirality, which is a crucial requirement for controlled current-induced dynamics. Our analysis reveals that the interplay between interlayer exchange and interlayer magnetic dipolar interactions plays a key role in significantly reducing the critical strength of the Dzyaloshinskii-Moriya interaction required to stabilize topological spin textures, such as AFM merons, making synthetic antiferromagnets a promising platform for next-generation spintronics applications.
\end{abstract}
\begin{document}

\flushbottom
\maketitle

\thispagestyle{empty}

\section*{Introduction}

The recent years have witnessed an increasing interest in chiral magnetic topological spin textures stabilized by the Dzyaloshinskii–Moriya interaction (DMI), such as skyrmions\cite{muhlbauer2009skyrmion, yu2010real}, biskyrmions\cite{yu2014biskyrmion, wang2016centrosymmetric, peng2017real}, hopfions\cite{kent2021creation,wang2019current}, chiral bobbers\cite{zheng2018experimental, ran2021creation}, and skyrmionic cocoons \cite{grelier2022three} due to their potential use as information carriers for high-density data storage and (un)conventional computing \cite{nagaosa2013topological, fert2013skyrmions, luo2018reconfigurable, raab2022brownian, zazvorka2019thermal}. For instance, skyrmions exhibit significant topological robustness \cite{je2020direct} and are amenable to electrical control \cite{jiang2015blowing, woo2016observation, litzius2017skyrmion, dohi2022thin }, but also entail disadvantages such as skyrmion Hall effect \cite{woo2016observation,litzius2017skyrmion,dohi2022thin}. The growing demand for high-speed, low-power technologies has therefore boosted the search for more complex topological spin textures beyond the skyrmion paradigm. Topological spin textures in in-plane magnets, such as merons and bimerons \cite{yu2018transformation, kharkov2017bound, janson2014quantum} are recently being explored, by virtue of their richer current-induced dynamics compared to skyrmions \cite{zarzuela2020stability} and the stackability property that allows for denser quasi-one-dimensional racetracks in three dimensions, resulting in higher storage density. Bimerons are robust topological textures that are homeomorphic to skyrmions and offer more topological states than conventional skyrmions, which makes them an important focus in fundamental quasi-particle research as well as topology-based computing approaches. Despite the advantage of stabilizing pure homochiral spin textures, ferromagnetic (FM) topological spin textures suffer from limitations in scalability with respect to sufficient thermal stability \cite{buttner2018theory}, stackability due to long-range magnetic dipolar interactions \cite{lemesh2019walker} and, gyrotropic forces resulting from their net intrinsic spin angular momentum.

Antiferromagnetic (AFM) systems can naturally overcome these inherent limitations of FM textures, due to their compensated spin angular momentum and negligible stray fields \cite{RevModPhys.90.015005, dohi2019formation, legrand2020room, dohi2022enhanced, caretta2018fast, hirata2019vanishing, jungwirth2016antiferromagnetic}. While one could envisage using single-crystalline AFMs, the inherent technological advantages are challenged by the difficulty of stabilizing pure homochiral spin textures. These challenges stem from the absence of significant Lifshitz invariants, resulting in the observed spin structures having random chirality \cite{ jani2021antiferromagnetic, PhysRevB.102.094415, amin2022antiferromagnetic, chmiel2018observation}. So far, the observation of in-plane (IP) topological spin textures such as bimerons has been limited in antiferromagnets to observing their helicity, in spite of recent advances in their creation via sophisticated protocols \cite{jani2021antiferromagnetic, PhysRevB.102.094415, amin2022antiferromagnetic, chmiel2018observation}. For the dynamics, the anticipated motion of topological spin textures in the presence of spin-orbit torques is heavily influenced by their helicity, leading to Bloch-type and Néel-type structures moving perpendicular and along the direction of spin-orbit torque (SOT), respectively \cite{dohi2022thin}. However, the lack of homochirality of spin textures in native single-crystalline antiferromagnets limits their use for controlled dynamics of spin textures and thus prevents their utilization in future spintronics devices.



An ideal system to explore and manipulate both structural and dynamical properties of (IP) topological spin textures are synthetic antiferromagnetic (SyAFM) platforms \cite{dohi2019formation,legrand2020room,dohi2022enhanced,juge2022skyrmions}, consisting of a multi-layered heterostructure made of FM thin films separated by nonmagnetic metallic spacers and antiferromagnetically coupled via the interlayer exchange interaction\cite{tomasello2017performance, chen2020realization}. By tailoring the amount of compensation they can exhibit an arbitrarily small magnetic moment and, therefore, combine the most interesting features of both FM and AFM scenarios: minimal stray fields, the ability to stabilize homochiral spin textures, and the potential for ultrafast spin dynamics; all in a device-compatible easy to fabricate polycrystalline multilayer setting. A precondition to assess the topological nature of such spin textures is to be able to measure their chirality. In this regard, SyAFMs offer the advantage to employ the advanced surface- or element-sensitive imaging methods available for FM. Nonetheless, these AFM IP topological spin textures can be formed locally during the magnetization reversal process\cite{kolesnikov2018composite}, spontaneously stable homochiral spin textures on a global scale are hitherto not observed.

In this article, we employ three-dimensional (3D) vector imaging of the staggered magnetization to demonstrate the successful stabilization of all memebers of the class of IP AFM topological spin textures emerging in a newly designed layered SyAFM, namely merons, antimerons, and bimerons at zero magnetic fields. Our experiments combine magnetic force microscopy (MFM), scanning electron microscopy with polarization analysis (SEMPA), and element-specific photoemission electron microscopy using the X-ray magnetic circular dichroism (XMCD-PEEM) that enable us to identify spin textures possessing enhanced stability, classified by an integer topological invariants and those that are topologically trivial. We find that in the vicinity of the spin-reorientation transition (SRT), where the effective anisotropy vanishes, a SyAFM platform can host homochiral AFM merons, as determined from their helicity and core polarity. Furthermore, their helicity can be easily tailored by the degree of magnetic compensation of the SyAFM, indicating thus that interlayer dipolar interactions play a significant role in the stabilization of these spin textures. Our micromagnetic and analytical calculations can fully explain the experimental observations, elucidate the mechanism for the stabilization of AFM topological textures in synthetic antiferromagnets, and describe the corresponding phase diagram. Our findings provide crucial insights into the formation and stability of homochiral IP AFM topological textures, which pave the way towards better scalable soliton-based technologies beyond the skyrmion paradigm.

\section*{Antiferromagnetic merons/antimerons in SyAFM platforms}

\begin{figure}[h!]
    \centering
    \includegraphics[width = 17.2cm]{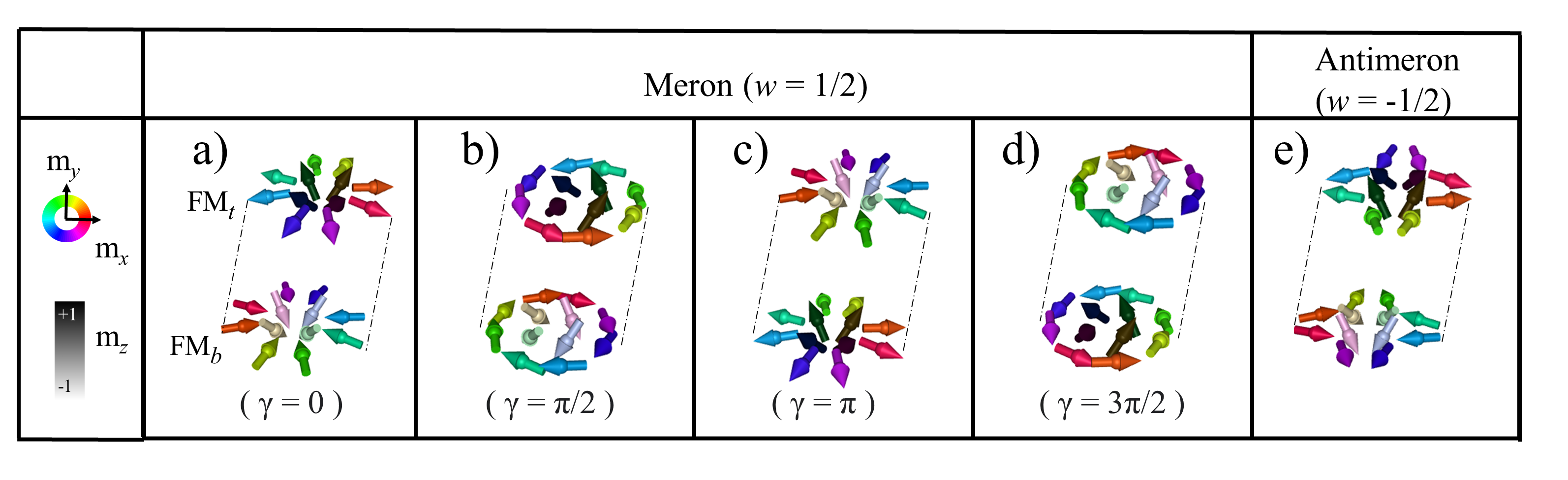}
    \caption{\textbf{Spin configuration of AFM merons and antimerons in a SyAFM platform} \textbf{(a)--(d)} AFM merons with helicities $\gamma = 0, \tfrac{\pi}{2},  \pi$, and $\tfrac{3\pi}{2}$, respectively. \textbf{(e)} AFM antimeron. Black and white arrows represent upward and downward core polarities, respectively, while the IP component of the moments is given by the color map in the top left corner. }
    \label{fig1}
\end{figure}

The magnetic order of the SyAFM platform can be described phenomenologically by two parameters that reflect both its AFM and FM character, depending on the chosen ratio of magnetic compensation: The Néel order parameter  $\bm{L}=\bm{M}_{\textrm{t}}-\bm{M}_{\textrm{b}}$ and the macroscopic spin density $\bm{M}=\bm{M}_{\textrm{t}}+\bm{M}_{\textrm{b}}$, where $\bm{M}_{\textrm{t}}$ and $\bm{M}_{\textrm{b}}$ denote the magnetization fields of the top and bottom FM layers in each double layer, respectively. As illustrated in Figure~\ref{fig1}, the topological spin textures in the SyAFM can be classified by their winding number $w$,  topological charge $Q$, and helicity $\gamma$, which are defined as follows: 1) $w$ provides a measure of the wrapping of the Néel order around the unit sphere and it reads $w=\pm1$ for skyrmions/bimerons, whereas it becomes $w=\tfrac{1}{2}$ for a meron and $w=-\tfrac{1}{2}$ for an antimeron. 2) $Q$ can be cast as the product of the winding number and the core polarity, namely $Q=w\cdot L_{z}|_{\textrm{core}}$, the core polarity being defined as the $z$ component of the Néel order at the texture core. 3) $\gamma$ is given, akin to skyrmions\cite{nagaosa2013topological}, by the angle between the IP projection of the Néel order and the radial direction. 
The helicity of these (anti)meron composites can be obtained from the IP rotation of the top-layer magnetization since the latter determines the direction of the Néel order. 

Néel-type merons are characterized by $\gamma=0$ or $\pi$, depending on the sign of the stabilizing DMI. Their spin structure is sketched in panels a) and c) of Figure~\ref{fig1}. Bloch-type merons are characterized by $\gamma=\tfrac{\pi}{2}$ or $\tfrac{3\pi}{2}$, as sketched in panels b) and d) of Figure~\ref{fig1}. Note that spin textures in adjacent FM layers exhibit identical winding numbers but opposite core polarities, so their helicities differ by a factor of $\pi$. 

\section*{Tuning the magnetic properties to stabilize AFM (anti)merons}

Choosing FM materials with low pinning, negligible perpendicular magnetic anisotropy (PMA) and finite DMI, as well as a strong AFM coupling between adjacent FM layers is a key requirement for the stabilization of (anti)merons and bimerons in SyAFM platforms as devised later in the discussion section.
\begin{figure*}[htb]
    \centering
    \includegraphics[width = 17.5cm]{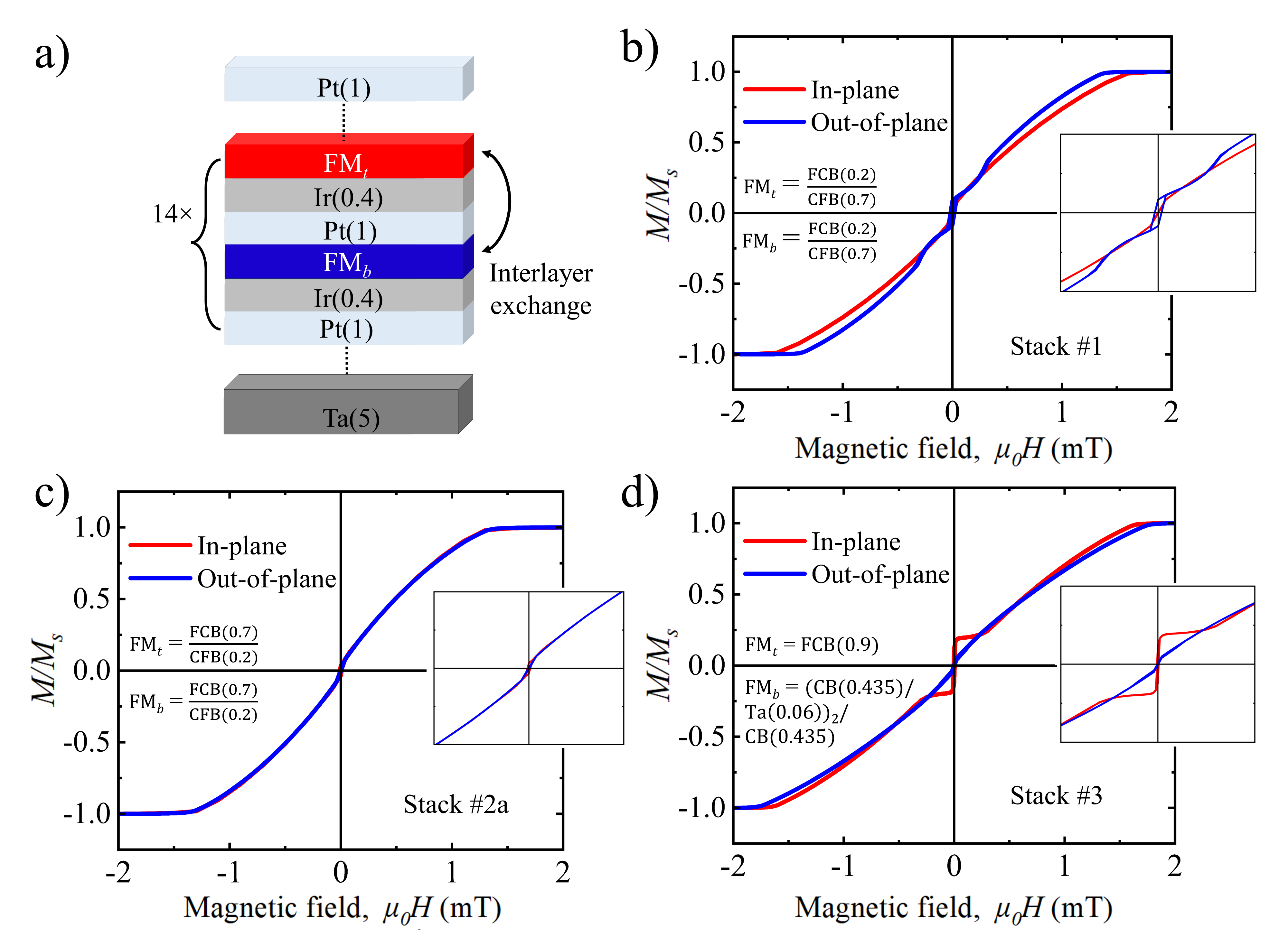}
   \caption{\textbf{Material stack and magnetic properties of the SyAFM.} \textbf{(a)} Multilayer structure for the SyAFM, where FM$_{\textrm{t}}$ and FM$_{\textrm{b}}$ denote the AFM-coupled top and bottom FM layers, respectively. The unit in parenthesis is in nm  \textbf{(b)--(d)} OOP (blue curve) and IP (red curve) hysteresis loops measured by means of SQUID magnetometry for the stack \textbf{(b)} \#1 (fully compensated OOP), \textbf{(c)} \#2a (fully compensated IP) and \textbf{(d)} \#3 (uncompensated IP). The fully compensated stack \#2a shows a complete overlap between the IP and OOP $M(H)$ curves, which indicates that the multilayer is in the vicinity of the SRT.}
    \label{fig2}
\end{figure*} 
We have therefore optimized a Pt/CoFeB/Ir-based multilayer SyAFM (see Methods for details). The FM films consist of the bilayer Fe$_{0.6}$Co$_{0.2}$B$_{0.2}$(FCB)/Co$_{0.6}$Fe$_{0.2}$B$_{0.2}$(CFB) and is sandwiched by nonmagnetic spacers made of a heavy metal bilayer (Pt and Ir). The latter breaks the mirror symmetry of the heterostructure and thus provides a finite DMI $D$. The top and bottom FM layers are denoted by FM$_{\textrm{t}}$ and FM$_{\textrm{b}}$, respectively, and their saturation magnetization by $M_{\textrm{s,t}}$ and $M_{\textrm{s,b}}$. The CFB layer induces PMA ($K_{\textrm{u}}$) at the interface with the heavy metal, whereas the thickness ratio to the FCB layer is used to control the magnetic dipolar anisotropy $K_{\textrm{d}}=-\tfrac{1}{2}\mu_0 M_{\textrm{s}}^2$.


Panels~\ref{fig2}(b)--(d) depict the $M(H)$ loops of the stacks $\#1$, $\#2$a and $\#3$, respectively, where the red (blue) curve represents the magnetic hysteresis loop for an IP (OOP) configuration of the external magnetic field. The stack $\#1$ has a small positive effective anisotropy. In Fig.~\ref{fig2}(c), red and blue hysteresis loops coincide, indicating a very small effective anisotropy $K_{\textrm{eff}}=-0.04$ MJ$\cdot$m$^{-2}$. Furthermore, the zero remanence makes the stack $\#2$a a potential candidate for hosting (bi)merons, as the formation of a multi-domain magnetic ground state is expected for this stack at zero fields. The effect of the magnetic compensation in synthetic antiferromagnets on the formation of meron structures has been studied via the stacks $\#2$b and $\#3$: the FM layers of the former have small (normalized) uncompensated magnetization $m_{\textrm{uncom}}=\tfrac{|M_{s,t}-M_{s,b}|}{M_{s,t}+M_{s,b}}=0.05$, which enables the detection of the OOP spin components of the meron textures via MFM imaging (see supplementary material (SM) section 4 for the $M(H)$ curve of stack $\#2$b). Figure~\ref{fig2}(d) shows the hysteresis loops of the stack $\#3$, which indicates a negative value for $K_{\textrm{eff}}$. 

\section*{3D-vector imaging of merons/antimerons in the SyAFM }
 \begin{figure*}[!ht]
    \centering
    \includegraphics[width = 17.6cm]{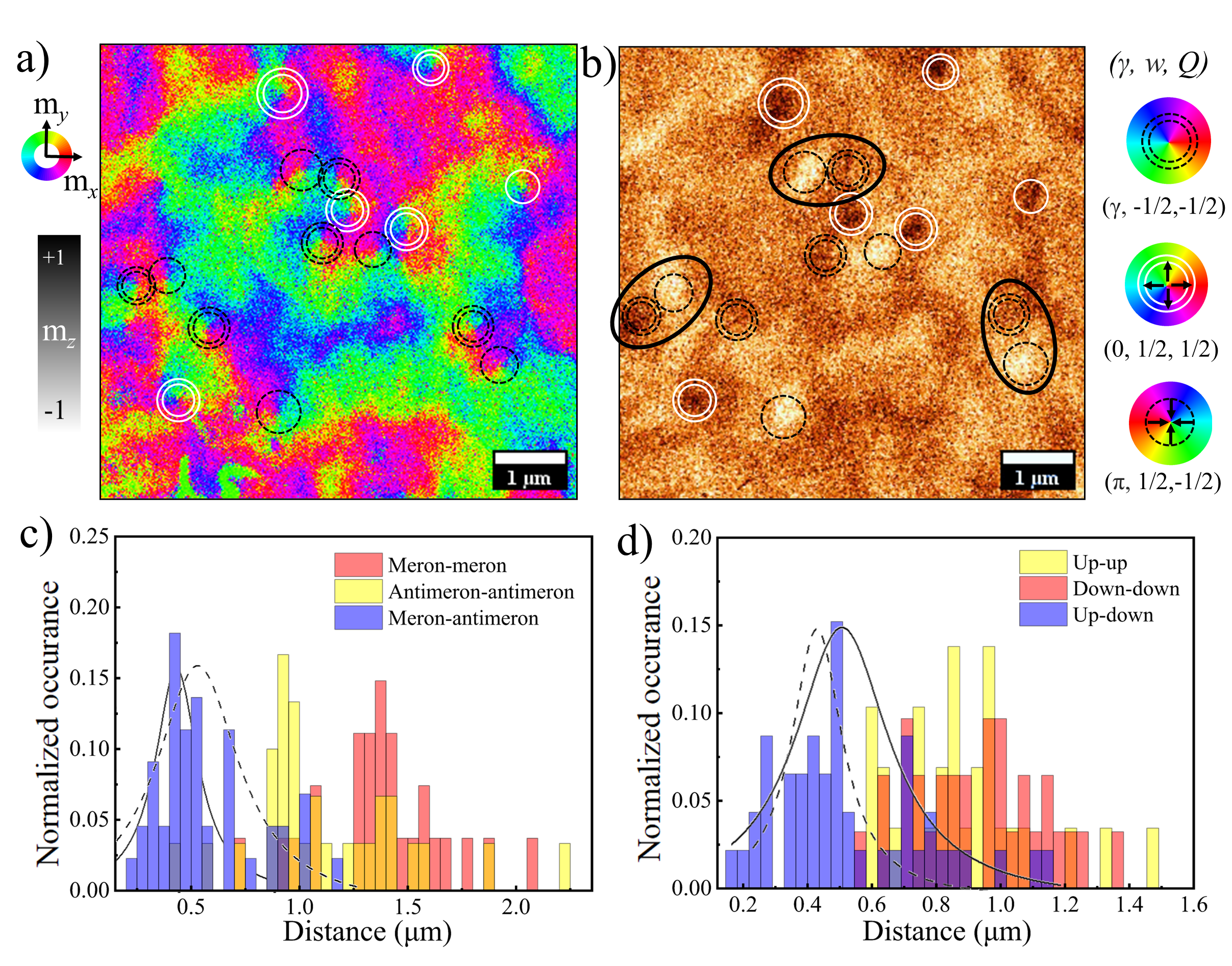}
    \caption{\textbf{Imaging the Néel order parameter of (anti)merons and bimerons in synthetic antiferromagnets.} \textbf{(a)} SEMPA image showing the IP spin components of the meron texture in the stack $\#2$b.
    \textbf{(b)} MFM image depicting the OOP spin component of the meron structure in the same area. Dark brown and white MFM contrasts indicate the upward and the downward direction, respectively. The color map for the SEMPA image is shown on the left side. Black dotted circles represent merons of helicity $\gamma = \pi$, whereas double black dotted circles indicate antimerons with $Q$ = $-\tfrac{1}{2}$. White circles represent merons having arbitrary helicity with $Q= \tfrac{1}{2}$ and white double circles denote merons of helicity $\gamma = 0$ with $Q$ = $\tfrac{1}{2}$. Two adjacent black circles (single and double) are identified as bimerons with net topological charge $Q = -1$ and additionally highlighted by ellipses. \textbf{(c)} Histogram of the next-neighbour separation between merons, antimerons and meron-antimeron composites obtained from SEMPA images. \textbf{(d)} Histogram of the next-neighbour separation between up-up, down-down, and up-down core polarities from MFM contrast.}
    \label{fig3}
\end{figure*}

Next we reveal the topological spin textures present and for the full vector reconstruction of the Néel order requires imaging of both IP and OOP spin components of the meron spin structues is required. Since the stack $\#2$a has full magnetic compensation, its OOP spin components are almost impossible to visualize via MFM. However, since the stacks $\#2$b and $\#3$ produce small stray fields, by using the combination of SEMPA and MFM imaging techniques at the same spot we can reconstruct the Néel order describing the topological textures in these SyAFM platforms.

Panels~\ref{fig3}(a) and~\ref{fig3}(b) show the SEMPA image (IP spin components) and the MFM image (OOP spin component) taken in the same area of the stack $\#2$b at room temperature and zero magnetic fields. The observed topological textures are created by applying a damped oscillating OOP magnetic field applied to imaging. White and dark brown contrasts in Figure~\ref{fig3}(b) show the downward and upward core polarity, respectively. Furthermore, black and white circles in both images correspond to $Q = -\tfrac{1}{2}$ and $Q=\tfrac{1}{2}$, respectively. We note that $M_{\textrm{s,t}}>M_{\textrm{s,b}}$, so the stray field detected in the MFM measurement is dominated by that of the topmost layer. The analysis of both images yields the presence of different topological spin textures: black dotted circles represent Néel-type merons with core polarity pointing downward ($L_{z}=-1$), namely the $\gamma = \pi$ merons from Fig.~\ref{fig1}(c). White double circles correspond to merons of helicity $\gamma = 0$ and upward core polarity, as described in panel~\ref{fig1}(a). We note that the helicities $\gamma = 0, \pi$ largely correspond to the core polarities $L_{z}=1$ and $-1$, respectively, which indicates the presence of homochiral merons in the system. Double black dotted circles mark antimerons with topological charge $Q=-\tfrac{1}{2}$, see Fig.~\ref{fig1}(e). Changes in $\gamma$ do not affect the topological structure of the antimeron except for a geometrical rotation of its IP spin components, so that we consider all of them to be topologically equivalent. The combination of a single black circle adjacent to a double one is identified as a bimeron with $Q = -1$ and marked by an ellipse. AFM coupling between the meronic spin textures present in the adjacent FM layers having  has been confirmed by means of XMCD-PEEM layer-resolved imaging (see SM section 2). 

By observing the arrangement shown in \ref{fig3}(a), it is evident that merons and antimerons are positioned in close proximity to each other. To gain a more comprehensive understanding of the range of their interaction, we conducted a statistical analysis of the distance between meronic textures over a larger sample area. Panel~\ref{fig3}(c) shows the histogram of separations between the constituents of various meron composites as seen in the SEMPA images. Meron-antimeron pairs average at a separation of (490 $\pm$ 32) nm, closer than that between two antimerons or two merons, suggesting a different interaction potential with an energy minimum at smaller distance between merons and antimerons due to the presence of DMI. Based on the core polarities, meron-antimeron composites can have non-zero topological charge or be trivial spin textures with $Q$ = 0. In panel~\ref{fig3}(d), histograms of the next-neighbour distances between core polarities as seen in MFM images are presented. The average separation between up-down core polarities, (425 $\pm$ 40) nm, coincides within the experimental uncertainty with the average separation between merons and antimerons in the SEMPA images and is smaller than the separation between up-up or down-down polarities. This supports the existence of non-zero topological charges in the meron-antimeron pairs and the dominance of bimerons in the sample as shown in Panels \ref{fig3}(a) and \ref{fig3}(b).

\section*{Tailoring the helicity of (anti)merons in synthetic antiferromagnets}
The measured helicity values are essential in identifying the mechanism of stabilization and in engineering the SOT-induced dynamics of meron structures.\begin{figure*}[!h]
    \centering
    \includegraphics[width = 17.5cm]{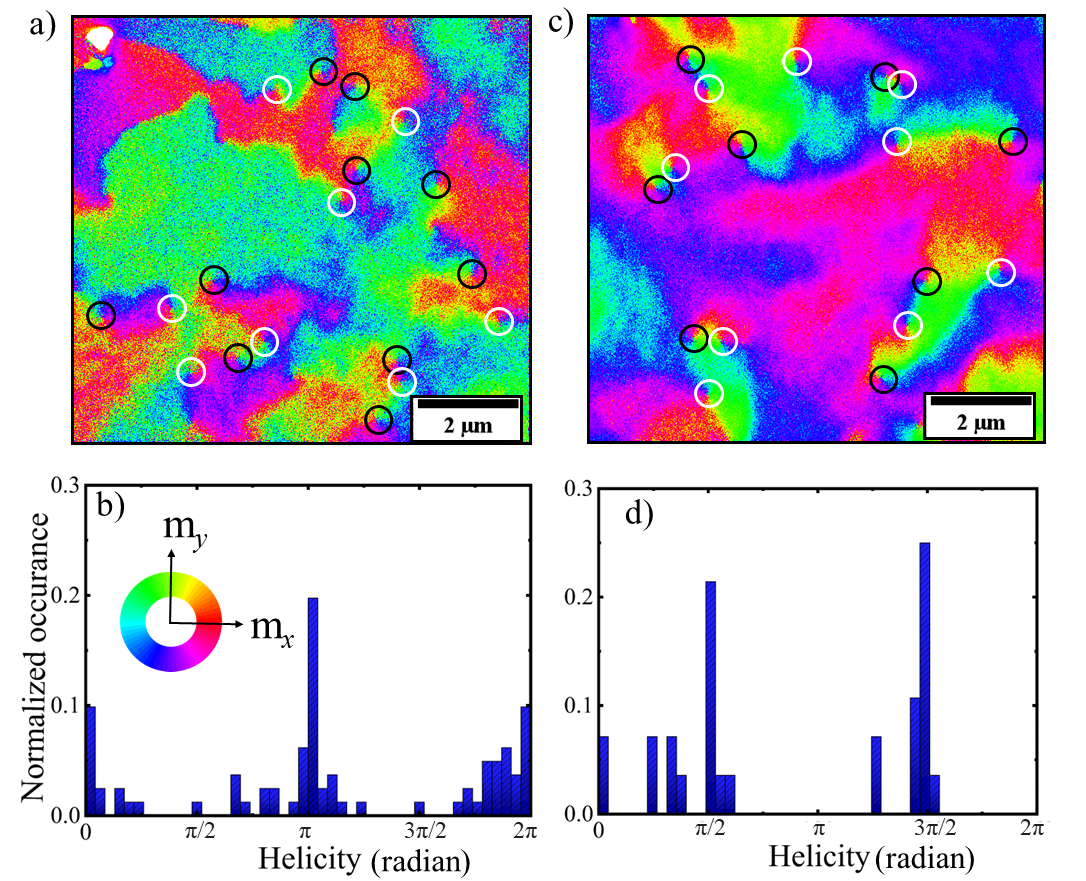}
    \caption{\textbf{Manipulating the helicity of (anti)merons in SyAFM}.
     \textbf{(a)} SEMPA image showing the IP spin components of the meronic spin texture in the stack $\#2$a indicating the IP orientation of the staggered magnetization. Black and white circles denote antimerons and merons, respectively. \textbf{b)} Distribution of helicities of the meron present in the SyAFM. The abudance of helicity at 0 and $\pi$ indicates homochiral Néel meron in the stack. \textbf{c)} SEMPA image showing the IP spin components of the meronic spin texture the (uncompensated) case of stack $\#3$. \textbf{(d)} shows the dominance of Bloch merons having helicity $\pi/2$ and 3$\pi/2$ in the stack. } 
    \label{fig4}
\end{figure*} In this section, we demonstrate the control of helicity in synthetic antiferromagnets through variation of the magnetic compensation ratio.

Figures~\ref{fig4}(a) shows a SEMPA images of the topmost-layer magnetization, hence the direction of the Néel order, for the stack $\#2$a in the absence of magnetic field. We have marked all merons with white circles and all antimerons with black circles, and find an almost equal proportion of both types of topological spin textures. We elucidate the relevance of the helicity by analyzing its values for the emergent merons through a histogram, as shown in panel~\ref{fig4}(b). This histogram has been generated by considering additional SEMPA images of the stack $\#2$a obtained under comparable conditions. Values of $\gamma=0,\pi$ are significantly favoured in this SyAFM platform, which corresponds to the Néel-type rotation, and therefore confirms that the stabilization mechanism for merons in the compensated case originates from the DMI \cite{zarzuela2020stability, moon2019existence,shen2020current}. Similarly, Néel bimerons are energetically favourable in the same range of DMI \cite{zarzuela2020stability}.

The more uncompensated case has been studied in stack $\#3$, which presents an uncompensated magnetization of $m_{\textrm{uncomp}}=0.20$ and thus a small but significant interlayer dipolar field. SEMPA images of its topmost layer magnetization is shown in figure ~\ref{fig4}(c) with the direction of the net IP magnetization. We observe again a nearly equal number of merons and antimerons (same white/black color convention as before). The analysis of the distribution of meron helicities, see panel~\ref{fig4}(d), yields a prevalence of values around $\gamma = \tfrac{\pi}{2}, \tfrac{3\pi}{2}$, which indicates a Bloch-type rotation. We conclude that the presence of a small uncompensated magnetization in the SyAFM promotes the stabilization of Bloch-meron textures. Thus by tuning the compensation ratio, we can effectively manipulate the helicity and consequently, the resulting SOT-induced dynamics.

\section*{Theoretical model and Discussion}

In the preceding sections, we experimentally demonstrate the occurrence of AFM merons, antimerons, and bimerons in the SyAFM platform. The stabilization of meronic spin textures in SyAFMs results from the subtle interplay between interlayer exchange, interlayer magnetic dipolar interactions (IMD), and interfacial DMI, as well as the effective anisotropies of the FM layers. This mechanism of bimeron stabilization in a SyAFM platform is analyzed and explained in this section, supported by theoretical models and micromagnetic simulations.

We start with a SyAFM that can be effectively described, irrespective of its magnetic compensation ratio, as a ferrimagnetic platform with magnetic sublattices given by the top and bottom FM layers, respectively. In the compensated case, the minimal model for the SyAFM contains exchange, DMI, and anisotropy contributions, and thus its total free energy reads
\begin{align}
\label{eq:E_functional}
\mathcal{E}[ \bm{L}]&=\int_{\mathcal{S}}d^{2}\bm{r}\,\bigg[\tfrac{A}{2} (\nabla\bm{L})^{2}  + D\bm{L} \cdot ( \tilde{\nabla} \times \boldsymbol{L} ) -K L_{z}^2\bigg],
\end{align}
where $A$ is the AFM spin stiffness constant and $\mathcal{S}$ denotes the SyAFM surface, see SM. The effective (easy-axis) anisotropy constant for the Néel order, $K=K_{\textrm{eff}}-\tfrac{H_{\textrm{d}}^{2}}{2\lambda\bm{L}^{2}}$, has an additional contribution originating in the competition between the interlayer exchange and IMD interactions. Here, $H_{\textrm{d}}$ and $\lambda$ denote the interlayer stray field and half of the interlayer exchange constant, respectively. In the vicinity of the SRT point, $K_{\textrm{eff}}\sim\tfrac{H_{\textrm{d}}^{2}}{2\lambda\bm{L}^{2}}\ll K_{\textrm{u}}$, and therefore the IMD field can induce the reorientation (from OOP to IP) of the staggered magnetization describing the SyAFM, which in turn favours the stability of in-plane meron textures. We note that FM systems can be tuned close to zero effective anisotropy but lack the IMD field, whereas AFM platforms have a spin-flop contribution to the effective anisotropy but their SRT is usually driven by temperature \cite{jani2021antiferromagnetic}. AFM lacks inversion symmetry due to antiferromagnetic ordering, which significantly reduces the magnitude of the Lifshitz invariants, and therefore it lacks the necessary criteria to stabilize homochiral spin structures. The interplay between the tunable (near-zero) effective FM-layer anisotropy and the IMD field lowers significantly the DMI needed to stabilize IP AFM textures, which makes synthetic antiferromagnets optimal platforms to explore the physics of these AFM spin textures. Furthermore, in the uncompensated scenario, a Zeeman-like interaction $-\tfrac{\Theta}{\bm{L}^{2}} H_{\textrm{d}}L_{z}$ contributes to the energetics of the SyAFM, which favours the OOP orientation of the Néel order and, therefore, the stability of Bloch-type IP merons at low DMI. Here, $\Theta=M_{s,t}^{2}-M_{s,b}^{2}$ parametrizes the absence of magnetic compensation in the SyAFM.
\begin{figure*}[!h]
    \centering
    \includegraphics[width = 16cm]{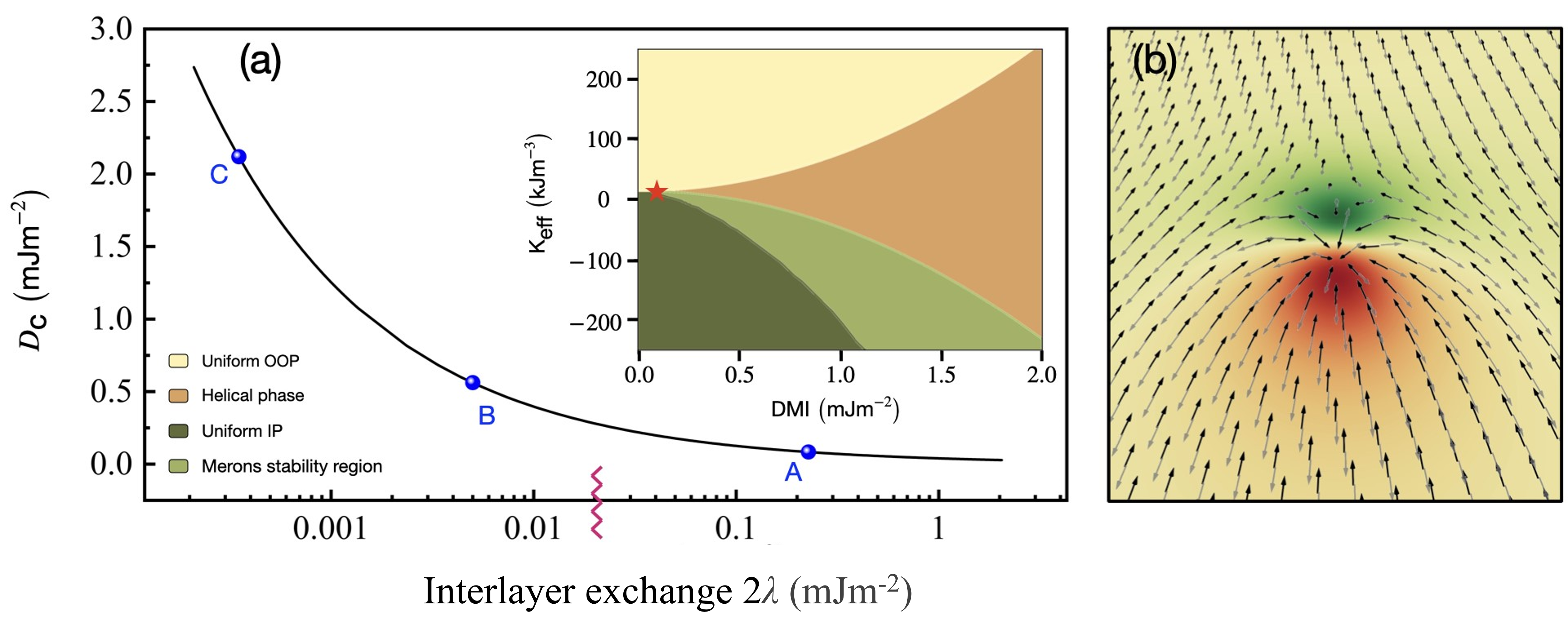}
    \caption{\textbf{Critical point and micromagnetics of bimerons in synthetic antiferromagnets.} \textbf{(a)} Dependence of the critical DMI $D_{\textrm{c}}$ on the interlayer exchange constant $2\lambda$. The inset shows the phase diagram corresponding to the point A along the curve $D_{\textrm{c}}(\lambda)$. The ground-state phases coalesce at the triple point $(\lambda, D_{\textrm{c}})$ in the $D-K_{\textrm{eff}}$ phase diagram, marked with a red star. Dark green, yellow and bronze colors illustrate the uniform IP, uniform OOP and helical in the $rz$ plane ground-state phases, respectively, with $r$ being any radial direction. Light green color depicts the region where AFM merons are stabilized in micromagnetic simulations. The magenta zigzag line illustrates the lower bound of optimal AFM interlayer exchange (energies) accessible in our stacks. The phase diagrams corresponding to the points B and C along the critical curve, which illustrate a displacement of the triple point to the right, are discussed in the SM section 5. \textbf{(b)} Illustration of an AFM bimeron corresponding to a system with parameters at point A. Black and grey arrows depict the antiparallel alignment of the spins of the top and bottom FM layers, respectively. Green and red colors show the OOP projections of the localized spins.}
    \label{fig5}
\end{figure*}

Panel~\ref{fig5}(a) depicts the dependence of the critical DMI ($D_\textrm{c}$) marking the onset of the meron instability (towards the helical phase) on the interlayer exchange constant $2\lambda$. For a strongly AFM-coupled SyAFM (i.e., large $\lambda$) in the vicinity of the SRT point (namely, $K_{\textrm{eff}}\simeq0$), only a very small DMI is needed to induce the phase transition from the uniform IP configuration to a helical phase (denoted by green and bronze domains in the inset). We note that the expression for the DMI at the triple point can be always obtained from the condition $K_{\textrm{eff}}=0$, which yields $D_{\textrm{c}} = \frac{4}{\pi} \sqrt{A \left[ \frac{H_{\textrm{d}}^2}{2 \lambda\bm{L}^{2}}  \right]}$ (see Methods). The low $D_{\textrm{c}}$ stems from the fact that the only contribution to the effective anisotropy at the SRT point comes from the IMD field. As the SyAFM is tuned away from the SRT point towards an IP configuration (i.e, $K<0$), the critical value $D_{\textrm{c}}$ increases since one needs to overcome a larger anisotropy barrier to induce the OOP tilting of the staggered magnetization. The phase boundary between the uniform IP and helical phases has been calculated analytically (see Methods) and described parametrically by the curve $D_{\textrm{c}}^{\textrm{IP}} = \frac{4}{\pi} \sqrt{A \left[\frac{H_{\textrm{d}}^2}{2 \lambda\bm{L}^{2}} -  K_{\textrm{eff}} \right]}$. Furthermore, the ground state is the uniform OOP configuration when $K>0$ (depicted as a yellow domain in the inset) and, as the DMI increases above $D_c$, a phase transition towards the helical state is induced, which is well known in magnetic films with PMA. We conclude the discussion by noting that, as seen in the inset C of SM figure s9, attaining the critical value $D_{\textrm{c}}\propto\tfrac{1}{\sqrt{\lambda}}$ for weak AFM interlayer couplings can be very difficult experimentally, consequently, systems with strong AFM interlayer coupling offer a more viable route for stabilizing bimerons.

\section*{Outlook}
In conclusion, we demonstrate the presence of chiral merons, antimerons, and topologically stabilized bimerons in synthetic antiferromagnets at zero magnetic fields. The direction of the net magnetization and the emergent field created by topology in bimerons are mutually orthogonal, a key difference to their out-of-plane counterparts, skyrmions. Thus meronic spin textures offer an approach to directly explore and tune the topological Hall physics. Their Hall signal will be directly sensitive to the topology, enabling the electrical readout of the topological winding number and leading to new possibilities for the designing of magnetic topology-based technology where the topology can encode the informations. Our findings show that these AFM textures can be detected with accurate helicity and topological charge through a combination of surface-sensitive SEMPA imaging and MFM imaging. The fully compensated synthetic antiferromagnets host homochiral Néel bimerons that are stable at room temperature. In SyAFM, bimerons exhibit an important advantage over previously demonstrated AFM bimerons due to the presence of homochiral spin textures, which makes them amenable to controlled manipulation using spin-orbit torques and in turn opens up new possibilities for designing spintronic devices in SyAFMs. This combine the advantages of FM, such as easy detection, with those of AFMs, such as the absence of the long-range stray fields.

\section*{Acknowledgments}
The authors thank A. Bose, A. Rajan and E. Galindez Ruales for their participation in additional experiments included in the Supplementary Material. This work has received funding from the European Union’s Horizon 2020 research and innovation program under the Marie Skłodowska-Curie Grant Agreement No. 860060 “Magnetism and the effect of Electric Field” (MagnEFi) as well as from Synergy Grant No. 856538, project “3D-MAGiC”. It has also been supported by the Deutsche Forschungsgemeinschaft (DFG, German Research Foundation) - TRR 173 -- 268565370 (project A03), the Grant Agency of the Czech Republic grant no. 19-28375X, and the Dynamics and Topology Centre TopDyn funded by the State of Rhineland Palatinate.

\section*{Methods}
\subsection*{Material deposition}
The thin-film material stacks were deposited on thermally oxidized Si/SiO2 substrates employing the Singulus Rotaris magnetron sputtering tool, which provides reproducibility and sub-Angstrom thickness accuracy. DC-magnetron sputtering at a base pressure of 4 $\times$ 10$^{-8}$ mbar was employed for the growth of the metallic layers Ta, Pt, Ir, Co$_{0.6}$Fe$_{0.2}$B$_{0.2}$(CFB), Fe$_{0.6}$Co$_{0.2}$B$_{0.2}$(FCB), and Co$_{0.8}$B$_{0.2}$ at room temperature. The respective deposition rates were determined with X-ray reflectivity to be 0.54, 0.91, 0.56, 0.51, 0.66, and 0.37 \AA s$^{-1}$ under pure Ar flow used as sputtering gas.

 The top Pt layer serves as capping for the material stack to prevent oxidation over time and during post processing (patterning). The thickness $d_{\textrm{Ir}}=0.4$ nm of the Ir layer is chosen so as to maximize the AFM interlayer exchange between the FM layers. The value of $x$ is 0.2 in stack $\#1$, whereas it is 0.7 in $\#2$a. In  $\#3$, the FM layers correspond to different values of the parameter $x$ ($x_{\textrm{FM}_{t}}=0.7$ and $x_{\textrm{FM}_{b}}=0.2$). We have kept the FM layers as thin as possible to maximize the interlayer exchange coupling ($d_{\textrm{FM}}=0.9$ nm) as well as optimized the ratio between the FCB ($x$ nm) and CFB ($0.9-x$ nm) thicknesses to be in the vicinity of the SRT (namely, to obtain a vanishing effective anisotropy $K_{\textrm{eff}}=K_{\textrm{u}}+K_{\textrm{d}}$).

As depicted in Figure~\ref{fig2}(a), a heterostructure consisting of multiple repetitions ($14$ times) of the single bilayer SyAFM has been prepared to 1) reduce the thermal diffusion of (anti)merons across the low-pinning SyAFM, and 2) obtain a high saturation field in the hysteresis loops. In Stack 3, bottom and top FM layers are made of Co$_{0.8}$B$_{0.2}$ and FCB, respectively, and have thicknesses of $d_{\textrm{FM}_{\textrm{b}}}=1.305$ nm and $d_{\textrm{FM}_{\textrm{t}}}=0.9$ nm, which yields the value $m_{\textrm{uncom}}=0.20$ for the uncompensated magnetization.

\subsection*{SEMPA imaging}
For imaging the in-plane component of the magnetic spin texture we used a surface-sensitive technique, the Scanning Electron Microscope with Polarization Analysis (SEMPA) \cite{schonke2018development}. SEMPA is a powerful in-house imaging technique that uses the spin-polarized secondary electrons emitted from a magnetic material and gives a two-dimensional (2D) vector map of the IP magnetization. The sensitivity of SEMPA is limited to $1-2$ nm depth from the surface, which enables us to image the topological spin textures present only in the topmost magnetic layer. This unique feature of SEMPA is especially effective on synthetic antiferromagnets enabling us to investigate the formation of topological spin textures even in a fully compensated composition. 

SEMPA color-coded images enable us to determine the winding number of the topological spin textures and classify them accordingly. Also, the sense of the in-plane rotation  gives information about the exact helicity of these meronic spin structures. We note that SEMPA images do not differentiate the OOP component of the magnetization, hence merons and antimerons of equal helicity (e.g., \ref{fig1} (a) and \ref{fig1} (f)) give similar SEMPA color contrast, as shown in figure \ref{fig1}(k). This prohibits to determine $Q_T$ solely from the IP contrast. Similarly, Figures \ref{fig1} (l)--(n) denote SEMPA images with $w$ = $\frac{1}{2}$ having $\gamma$  values $\tfrac{\pi}{2}$, $\pi$, and $\tfrac{3\pi}{2}$.

\subsection*{Micromagnetic approach}

\subsubsection*{Analytical expression for the phase boundaries}

We explore the possible ground states of the model~\eqref{eq:E_functional} along the lines of Ref. \cite{zarzuela2020stability}. We consider the most generic ansatz for a helix in the real space, which can be parametrized by the normal $\vec{\bm{n}}$ to the plane of the helix and the helical pitch vector $\vec{q}$. The Néel order can be cast in terms of this parametrization as
\begin{align}
\bm{l}(\vec{r}\,)=\cos(\vec{q}\cdot\vec{r})\hat{\bm{e}}_{1}+\sin(\vec{q}\cdot\vec{r})\hat{\bm{e}}_{2}+m_{0}\hat{\bm{n}} 
\label{Eq_L_in_spin_space}
\end{align}
where $\bm{l}(\vec{r}\,) = \bm{L}(\vec{r}\,)/|\bm{L}|$ and $\{\hat{\bm{e}}_{1},\hat{\bm{e}}_{2},\vec{\bm{n}}\}$ defines a local frame of reference in the spin space. Upon substituting this expression into Eq.~\eqref{eq:E_functional}, we obtain the following identity for the energy density functional:
\begin{align}
\label{Eq_Energy_helical_ansatz}
\varepsilon\big[\bm{l}(\vec{r}\hspace{0.05cm})\big]&=
\tfrac{1}{1+m_{0}^{2}} \Big\{
\tfrac{J}{2} \vec{q}\hspace{0.07cm}^{2}
+ D \left(q_{x}\sin\theta\sin\phi-q_{y}\sin\theta\cos\phi\right) 
+ K \big( \left[ \tfrac{1}{2}+m_{0}^{2} \right] + \left[ \tfrac{1}{2}-m_{0}^{2} \right] \cos^{2}\theta \big)
\Big\}.
\end{align}

This functional is minimized with respect to the variables $\{\theta,\phi,\vec{q},m_{0}\}$ and the different possible extrema are found. The lowest energy configuration for a given set of parameters determines the ground state. The phase boundaries separating any two possible ground states are determined by equating their corresponding energies, from which the expression of the DMI $D$ as a function of $K_\textrm{eff}$ is obtained. For instance, the phase boundary for the OOP-helical transition is parametrized by the curve $D_{\textrm{c}}^{\textrm{OOP}}(\lambda) = \frac{4}{\pi} \sqrt{A \left[ K_{\textrm{eff}}  - \frac{H_{\textrm{d}}^2}{2 \lambda\bm{L}^{2}}  \right]}$.

\subsubsection*{Micromagnetic simulations}
Micromagnetic simulations were performed using the Mumax3 software \cite{de2017modelling}. The following setup was implemented in the simulations leading to Fig.~\ref{fig5}.  A bilayer square geometry of lateral size 256 nm and thickness 1 nm for each of the FM layers was considered and dipolar interaction included. The system was discretized with a mesh size of $1  \times 1  \times 1$ nm$^{3}$ and periodic boundary conditions along the $x$ and $y$ directions were imposed, with period equal to 16 repetitions. The material parameters are  $A = 1  \times 10^{-11} $ Jm$^{-1}$ for the exchange constant,  $M_s$ = 0.145 MAm$^{-1}$  for the saturation magnetization and $\alpha = 0.01 $ for the Gilbert damping. The strength of the interlayer exchange coupling was chosen to be $\lambda = 0.44  \times 10^{-3} $ Jm$^{-2} $, which corresponds to the value obtained from SQUID measurements. We note that, in Mumax$^{3}$, interlayer exchange interactions are properly accounted for by rescaling the material parameters by the thickness of the spacer (see ext$\_$scaleExchange function)\cite{de2017modelling}. To explore the $D-K_{\textrm{eff}}$ phase diagram, the effective out-of-plane uniaxial anisotropy $K_{\textrm{eff}}$ and the DMI $D$ were varied in the range $\left[ -3 \times 10^{5}, 3 \times 10^{5} \right]$ Jm$^{-3} $ and $\left[ 0, 2 \times 10^{-3} \right]$ Jm$^{-2} $, respectively. An initial meron configuration is chosen in the simulations, which is minimized to find the equilibrium configuration. The parameter space of  $(D,K_{\textrm{eff}} )$ was swept to obtain the light green shaded region in Fig.~\ref{fig5}.

The stability of the bimerons was confirmed via micromagnetic simulations performed on MuMax$^3$. Their size and shape were analyzed in the domain $D<D_{\textrm{c}}^{\textrm{IP}}$ of the parameter space $\lambda-D$. Panel~\ref{fig5}(b) depicts the magnetization profile of a bimeron stabilized for the material parameters summarised in the Methods section. To gain deeper insight into the impact of various magnetic parameters on the properties of AFM bimerons, we conduct micromagnetic simulations (see SM section 6). Our simulations demonstrate that both the DMI and the $K_z$ jointly assist in the formation of larger bimerons in SyAFM, while stronger interlayer exchange stabilizes smaller bimerons. These findings show that by tunning the properties we can design and optimize of AFM bimeron-based devices.

\newpage
\bibstyle{nature}
\bibliography{getwriting}

\begin{thebibliography}{10}
\urlstyle{rm}
\expandafter\ifx\csname url\endcsname\relax
  \def\url#1{\texttt{#1}}\fi
\expandafter\ifx\csname urlprefix\endcsname\relax\def\urlprefix{URL }\fi
\expandafter\ifx\csname doiprefix\endcsname\relax\def\doiprefix{DOI: }\fi
\providecommand{\bibinfo}[2]{#2}
\providecommand{\eprint}[2][]{\url{#2}}

\bibitem{muhlbauer2009skyrmion}
\bibinfo{author}{M{\"u}hlbauer, S.} \emph{et~al.}
\newblock \bibinfo{journal}{\bibinfo{title}{Skyrmion lattice in a chiral
  magnet}}.
\newblock {\emph{\JournalTitle{Science}}} \textbf{\bibinfo{volume}{323}},
  \bibinfo{pages}{915--919} (\bibinfo{year}{2009}).

\bibitem{yu2010real}
\bibinfo{author}{Yu, X.} \emph{et~al.}
\newblock \bibinfo{journal}{\bibinfo{title}{Real-space observation of a
  two-dimensional skyrmion crystal}}.
\newblock {\emph{\JournalTitle{Nature}}} \textbf{\bibinfo{volume}{465}},
  \bibinfo{pages}{901--904} (\bibinfo{year}{2010}).

\bibitem{yu2014biskyrmion}
\bibinfo{author}{Yu, X.} \emph{et~al.}
\newblock \bibinfo{journal}{\bibinfo{title}{Biskyrmion states and their
  current-driven motion in a layered manganite}}.
\newblock {\emph{\JournalTitle{Nature Communications}}}
  \textbf{\bibinfo{volume}{5}}, \bibinfo{pages}{1--7} (\bibinfo{year}{2014}).

\bibitem{wang2016centrosymmetric}
\bibinfo{author}{Wang, W.} \emph{et~al.}
\newblock \bibinfo{journal}{\bibinfo{title}{A centrosymmetric hexagonal magnet
  with superstable biskyrmion magnetic nanodomains in a wide temperature range
  of 100--340 k}}.
\newblock {\emph{\JournalTitle{Advanced Materials}}}
  \textbf{\bibinfo{volume}{28}}, \bibinfo{pages}{6887--6893}
  (\bibinfo{year}{2016}).

\bibitem{peng2017real}
\bibinfo{author}{Peng, L.} \emph{et~al.}
\newblock \bibinfo{journal}{\bibinfo{title}{Real-space observation of
  nonvolatile zero-field biskyrmion lattice generation in mnniga magnet}}.
\newblock {\emph{\JournalTitle{Nano Letters}}} \textbf{\bibinfo{volume}{17}},
  \bibinfo{pages}{7075--7079} (\bibinfo{year}{2017}).

\bibitem{kent2021creation}
\bibinfo{author}{Kent, N.} \emph{et~al.}
\newblock \bibinfo{journal}{\bibinfo{title}{Creation and observation of
  hopfions in magnetic multilayer systems}}.
\newblock {\emph{\JournalTitle{Nature Communications}}}
  \textbf{\bibinfo{volume}{12}}, \bibinfo{pages}{1--7} (\bibinfo{year}{2021}).

\bibitem{wang2019current}
\bibinfo{author}{Wang, X.}, \bibinfo{author}{Qaiumzadeh, A.} \&
  \bibinfo{author}{Brataas, A.}
\newblock \bibinfo{journal}{\bibinfo{title}{Current-driven dynamics of magnetic
  hopfions}}.
\newblock {\emph{\JournalTitle{Physical Review Letters}}}
  \textbf{\bibinfo{volume}{123}}, \bibinfo{pages}{147203}
  (\bibinfo{year}{2019}).

\bibitem{zheng2018experimental}
\bibinfo{author}{Zheng, F.} \emph{et~al.}
\newblock \bibinfo{journal}{\bibinfo{title}{Experimental observation of chiral
  magnetic bobbers in b20-type fege}}.
\newblock {\emph{\JournalTitle{Nature Nanotechnology}}}
  \textbf{\bibinfo{volume}{13}}, \bibinfo{pages}{451--455}
  (\bibinfo{year}{2018}).

\bibitem{ran2021creation}
\bibinfo{author}{Ran, K.} \emph{et~al.}
\newblock \bibinfo{journal}{\bibinfo{title}{Creation of a chiral bobber lattice
  in helimagnet-multilayer heterostructures}}.
\newblock {\emph{\JournalTitle{Physical Review Letters}}}
  \textbf{\bibinfo{volume}{126}}, \bibinfo{pages}{017204}
  (\bibinfo{year}{2021}).

\bibitem{grelier2022three}
\bibinfo{author}{Grelier, M.} \emph{et~al.}
\newblock \bibinfo{journal}{\bibinfo{title}{Three-dimensional skyrmionic
  cocoons in magnetic multilayers}}.
\newblock {\emph{\JournalTitle{arXiv:2205.01172}}}  (\bibinfo{year}{2022}).

\bibitem{nagaosa2013topological}
\bibinfo{author}{Nagaosa, N.} \& \bibinfo{author}{Tokura, Y.}
\newblock \bibinfo{journal}{\bibinfo{title}{Topological properties and dynamics
  of magnetic skyrmions}}.
\newblock {\emph{\JournalTitle{Nature Nanotechnology}}}
  \textbf{\bibinfo{volume}{8}}, \bibinfo{pages}{899} (\bibinfo{year}{2013}).

\bibitem{fert2013skyrmions}
\bibinfo{author}{Fert, A.}, \bibinfo{author}{Cros, V.} \&
  \bibinfo{author}{Sampaio, J.}
\newblock \bibinfo{journal}{\bibinfo{title}{Skyrmions on the track}}.
\newblock {\emph{\JournalTitle{Nature Nanotechnology}}}
  \textbf{\bibinfo{volume}{8}}, \bibinfo{pages}{152--156}
  (\bibinfo{year}{2013}).

\bibitem{luo2018reconfigurable}
\bibinfo{author}{Luo, S.} \emph{et~al.}
\newblock \bibinfo{journal}{\bibinfo{title}{Reconfigurable skyrmion logic
  gates}}.
\newblock {\emph{\JournalTitle{Nano Letters}}} \textbf{\bibinfo{volume}{18}},
  \bibinfo{pages}{1180--1184} (\bibinfo{year}{2018}).

\bibitem{raab2022brownian}
\bibinfo{author}{Raab, K.} \emph{et~al.}
\newblock \bibinfo{journal}{\bibinfo{title}{Brownian reservoir computing
  realized using geometrically confined skyrmion dynamics}}.
\newblock {\emph{\JournalTitle{Nature Communications}}}
  \textbf{\bibinfo{volume}{13}}, \bibinfo{pages}{1--6} (\bibinfo{year}{2022}).

\bibitem{zazvorka2019thermal}
\bibinfo{author}{Z{\'a}zvorka, J.} \emph{et~al.}
\newblock \bibinfo{journal}{\bibinfo{title}{Thermal skyrmion diffusion used in
  a reshuffler device}}.
\newblock {\emph{\JournalTitle{Nature Nanotechnology}}}
  \textbf{\bibinfo{volume}{14}}, \bibinfo{pages}{658--661}
  (\bibinfo{year}{2019}).

\bibitem{je2020direct}
\bibinfo{author}{Je, S.-G.} \emph{et~al.}
\newblock \bibinfo{journal}{\bibinfo{title}{Direct demonstration of topological
  stability of magnetic skyrmions via topology manipulation}}.
\newblock {\emph{\JournalTitle{ACS nano}}} \textbf{\bibinfo{volume}{14}},
  \bibinfo{pages}{3251--3258} (\bibinfo{year}{2020}).

\bibitem{jiang2015blowing}
\bibinfo{author}{Jiang, W.} \emph{et~al.}
\newblock \bibinfo{journal}{\bibinfo{title}{Blowing magnetic skyrmion
  bubbles}}.
\newblock {\emph{\JournalTitle{Science}}} \textbf{\bibinfo{volume}{349}},
  \bibinfo{pages}{283--286} (\bibinfo{year}{2015}).

\bibitem{woo2016observation}
\bibinfo{author}{Woo, S.} \emph{et~al.}
\newblock \bibinfo{journal}{\bibinfo{title}{Observation of room-temperature
  magnetic skyrmions and their current-driven dynamics in ultrathin metallic
  ferromagnets}}.
\newblock {\emph{\JournalTitle{Nature Materials}}}
  \textbf{\bibinfo{volume}{15}}, \bibinfo{pages}{501--506}
  (\bibinfo{year}{2016}).

\bibitem{litzius2017skyrmion}
\bibinfo{author}{Litzius, K.} \emph{et~al.}
\newblock \bibinfo{journal}{\bibinfo{title}{Skyrmion hall effect revealed by
  direct time-resolved x-ray microscopy}}.
\newblock {\emph{\JournalTitle{Nature Physics}}} \textbf{\bibinfo{volume}{13}},
  \bibinfo{pages}{170} (\bibinfo{year}{2017}).

\bibitem{dohi2022thin}
\bibinfo{author}{Dohi, T.}, \bibinfo{author}{Reeve, R.~M.} \&
  \bibinfo{author}{Kl{\"a}ui, M.}
\newblock \bibinfo{journal}{\bibinfo{title}{Thin film skyrmionics}}.
\newblock {\emph{\JournalTitle{Annual Review of Condensed Matter Physics}}}
  \textbf{\bibinfo{volume}{13}}, \bibinfo{pages}{73--95}
  (\bibinfo{year}{2022}).

\bibitem{yu2018transformation}
\bibinfo{author}{Yu, X.} \emph{et~al.}
\newblock \bibinfo{journal}{\bibinfo{title}{Transformation between meron and
  skyrmion topological spin textures in a chiral magnet}}.
\newblock {\emph{\JournalTitle{Nature}}} \textbf{\bibinfo{volume}{564}},
  \bibinfo{pages}{95--98} (\bibinfo{year}{2018}).

\bibitem{kharkov2017bound}
\bibinfo{author}{Kharkov, Y.}, \bibinfo{author}{Sushkov, O.} \&
  \bibinfo{author}{Mostovoy, M.}
\newblock \bibinfo{journal}{\bibinfo{title}{Bound states of skyrmions and
  merons near the lifshitz point}}.
\newblock {\emph{\JournalTitle{Physical Review Letters}}}
  \textbf{\bibinfo{volume}{119}}, \bibinfo{pages}{207201}
  (\bibinfo{year}{2017}).

\bibitem{janson2014quantum}
\bibinfo{author}{Janson, O.} \emph{et~al.}
\newblock \bibinfo{journal}{\bibinfo{title}{The quantum nature of skyrmions and
  half-skyrmions in \ch{Cu2OSeO3}}}.
\newblock {\emph{\JournalTitle{Nature Communications}}}
  \textbf{\bibinfo{volume}{5}}, \bibinfo{pages}{1--11} (\bibinfo{year}{2014}).

\bibitem{zarzuela2020stability}
\bibinfo{author}{Zarzuela, R.}, \bibinfo{author}{Bharadwaj, V.~K.},
  \bibinfo{author}{Kim, K.-W.}, \bibinfo{author}{Sinova, J.} \&
  \bibinfo{author}{Everschor-Sitte, K.}
\newblock \bibinfo{journal}{\bibinfo{title}{Stability and dynamics of in-plane
  skyrmions in collinear ferromagnets}}.
\newblock {\emph{\JournalTitle{Physical Review B}}}
  \textbf{\bibinfo{volume}{101}}, \bibinfo{pages}{054405}
  (\bibinfo{year}{2020}).

\bibitem{buttner2018theory}
\bibinfo{author}{B{\"u}ttner, F.}, \bibinfo{author}{Lemesh, I.} \&
  \bibinfo{author}{Beach, G.~S.}
\newblock \bibinfo{journal}{\bibinfo{title}{Theory of isolated magnetic
  skyrmions: From fundamentals to room temperature applications}}.
\newblock {\emph{\JournalTitle{Sci. Rep.}}} \textbf{\bibinfo{volume}{8}},
  \bibinfo{pages}{4464} (\bibinfo{year}{2018}).

\bibitem{lemesh2019walker}
\bibinfo{author}{Lemesh, I.} \& \bibinfo{author}{Beach, G.~S.}
\newblock \bibinfo{journal}{\bibinfo{title}{Walker breakdown with a twist:
  Dynamics of multilayer domain walls and skyrmions driven by spin-orbit
  torque}}.
\newblock {\emph{\JournalTitle{Physical Review Applied}}}
  \textbf{\bibinfo{volume}{12}}, \bibinfo{pages}{044031}
  (\bibinfo{year}{2019}).

\bibitem{RevModPhys.90.015005}
\bibinfo{author}{Baltz, V.} \emph{et~al.}
\newblock \bibinfo{journal}{\bibinfo{title}{Antiferromagnetic spintronics}}.
\newblock {\emph{\JournalTitle{Rev. Mod. Phys.}}}
  \textbf{\bibinfo{volume}{90}}, \bibinfo{pages}{015005},
  \doiprefix\url{10.1103/RevModPhys.90.015005} (\bibinfo{year}{2018}).

\bibitem{dohi2019formation}
\bibinfo{author}{Dohi, T.}, \bibinfo{author}{DuttaGupta, S.},
  \bibinfo{author}{Fukami, S.} \& \bibinfo{author}{Ohno, H.}
\newblock \bibinfo{journal}{\bibinfo{title}{Formation and current-induced
  motion of synthetic antiferromagnetic skyrmion bubbles}}.
\newblock {\emph{\JournalTitle{Nature Communications}}}
  \textbf{\bibinfo{volume}{10}}, \bibinfo{pages}{1--6} (\bibinfo{year}{2019}).

\bibitem{legrand2020room}
\bibinfo{author}{Legrand, W.} \emph{et~al.}
\newblock \bibinfo{journal}{\bibinfo{title}{Room-temperature stabilization of
  antiferromagnetic skyrmions in synthetic antiferromagnets}}.
\newblock {\emph{\JournalTitle{Nature Materials}}}
  \textbf{\bibinfo{volume}{19}}, \bibinfo{pages}{34--42}
  (\bibinfo{year}{2020}).

\bibitem{dohi2022enhanced}
\bibinfo{author}{Dohi, T.} \emph{et~al.}
\newblock \bibinfo{journal}{\bibinfo{title}{Enhanced thermally-activated
  skyrmion diffusion in synthetic antiferromagnetic systems with tunable
  effective topological charge}}.
\newblock {\emph{\JournalTitle{arXiv preprint arXiv:2206.00791}}}
  (\bibinfo{year}{2022}).

\bibitem{caretta2018fast}
\bibinfo{author}{Caretta, L.} \emph{et~al.}
\newblock \bibinfo{journal}{\bibinfo{title}{Fast current-driven domain walls
  and small skyrmions in a compensated ferrimagnet}}.
\newblock {\emph{\JournalTitle{Nature Nanotechnology}}}
  \textbf{\bibinfo{volume}{13}}, \bibinfo{pages}{1154--1160}
  (\bibinfo{year}{2018}).

\bibitem{hirata2019vanishing}
\bibinfo{author}{Hirata, Y.} \emph{et~al.}
\newblock \bibinfo{journal}{\bibinfo{title}{Vanishing skyrmion hall effect at
  the angular momentum compensation temperature of a ferrimagnet}}.
\newblock {\emph{\JournalTitle{Nature Nanotechnology}}}
  \textbf{\bibinfo{volume}{14}}, \bibinfo{pages}{232--236}
  (\bibinfo{year}{2019}).

\bibitem{jungwirth2016antiferromagnetic}
\bibinfo{author}{Jungwirth, T.}, \bibinfo{author}{Marti, X.},
  \bibinfo{author}{Wadley, P.} \& \bibinfo{author}{Wunderlich, J.}
\newblock \bibinfo{journal}{\bibinfo{title}{Antiferromagnetic spintronics}}.
\newblock {\emph{\JournalTitle{Nature Nanotechnology}}}
  \textbf{\bibinfo{volume}{11}}, \bibinfo{pages}{231--241}
  (\bibinfo{year}{2016}).

\bibitem{jani2021antiferromagnetic}
\bibinfo{author}{Jani, H.} \emph{et~al.}
\newblock \bibinfo{journal}{\bibinfo{title}{Antiferromagnetic half-skyrmions
  and bimerons at room temperature}}.
\newblock {\emph{\JournalTitle{Nature}}} \textbf{\bibinfo{volume}{590}},
  \bibinfo{pages}{74--79} (\bibinfo{year}{2021}).

\bibitem{PhysRevB.102.094415}
\bibinfo{author}{Ross, A.} \emph{et~al.}
\newblock \bibinfo{journal}{\bibinfo{title}{Structural sensitivity of the spin
  hall magnetoresistance in antiferromagnetic thin films}}.
\newblock {\emph{\JournalTitle{Phys. Rev. B}}} \textbf{\bibinfo{volume}{102}},
  \bibinfo{pages}{094415} (\bibinfo{year}{2020}).

\bibitem{amin2022antiferromagnetic}
\bibinfo{author}{Amin, O.} \emph{et~al.}
\newblock \bibinfo{journal}{\bibinfo{title}{Antiferromagnetic half-skyrmions
  electrically generated and controlled at room temperature}}.
\newblock {\emph{\JournalTitle{arXiv preprint arXiv:2207.00286}}}
  (\bibinfo{year}{2022}).

\bibitem{chmiel2018observation}
\bibinfo{author}{Chmiel, F.~P.} \emph{et~al.}
\newblock \bibinfo{journal}{\bibinfo{title}{Observation of magnetic vortex
  pairs at room temperature in a planar $\alpha$-fe2o3/co heterostructure}}.
\newblock {\emph{\JournalTitle{Nature Materials}}}
  \textbf{\bibinfo{volume}{17}}, \bibinfo{pages}{581--585}
  (\bibinfo{year}{2018}).

\bibitem{juge2022skyrmions}
\bibinfo{author}{Juge, R.} \emph{et~al.}
\newblock \bibinfo{journal}{\bibinfo{title}{Skyrmions in synthetic
  antiferromagnets and their nucleation via electrical current and ultra-fast
  laser illumination}}.
\newblock {\emph{\JournalTitle{Nature Communications}}}
  \textbf{\bibinfo{volume}{13}}, \bibinfo{pages}{4807} (\bibinfo{year}{2022}).

\bibitem{tomasello2017performance}
\bibinfo{author}{Tomasello, R.} \emph{et~al.}
\newblock \bibinfo{journal}{\bibinfo{title}{Performance of synthetic
  antiferromagnetic racetrack memory: domain wall versus skyrmion}}.
\newblock {\emph{\JournalTitle{Journal of Physics D: Applied Physics}}}
  \textbf{\bibinfo{volume}{50}}, \bibinfo{pages}{325302}
  (\bibinfo{year}{2017}).

\bibitem{chen2020realization}
\bibinfo{author}{Chen, R.} \emph{et~al.}
\newblock \bibinfo{journal}{\bibinfo{title}{Realization of isolated and
  high-density skyrmions at room temperature in uncompensated synthetic
  antiferromagnets}}.
\newblock {\emph{\JournalTitle{Nano Letters}}} \textbf{\bibinfo{volume}{20}},
  \bibinfo{pages}{3299--3305} (\bibinfo{year}{2020}).

\bibitem{kolesnikov2018composite}
\bibinfo{author}{Kolesnikov, A.} \emph{et~al.}
\newblock \bibinfo{journal}{\bibinfo{title}{Composite topological structure of
  domain walls in synthetic antiferromagnets}}.
\newblock {\emph{\JournalTitle{Sci. Rep.}}} \textbf{\bibinfo{volume}{8}},
  \bibinfo{pages}{1--9} (\bibinfo{year}{2018}).

\bibitem{moon2019existence}
\bibinfo{author}{Moon, K.-W.}, \bibinfo{author}{Yoon, J.},
  \bibinfo{author}{Kim, C.} \& \bibinfo{author}{Hwang, C.}
\newblock \bibinfo{journal}{\bibinfo{title}{Existence of in-plane magnetic
  skyrmion and its motion under current flow}}.
\newblock {\emph{\JournalTitle{Physical Review Applied}}}
  \textbf{\bibinfo{volume}{12}}, \bibinfo{pages}{064054}
  (\bibinfo{year}{2019}).

\bibitem{shen2020current}
\bibinfo{author}{Shen, L.} \emph{et~al.}
\newblock \bibinfo{journal}{\bibinfo{title}{Current-induced dynamics and chaos
  of antiferromagnetic bimerons}}.
\newblock {\emph{\JournalTitle{Physical Review Letters}}}
  \textbf{\bibinfo{volume}{124}}, \bibinfo{pages}{037202}
  (\bibinfo{year}{2020}).

\bibitem{schonke2018development}
\bibinfo{author}{Sch{\"o}nke, D.}, \bibinfo{author}{Oelsner, A.},
  \bibinfo{author}{Krautscheid, P.}, \bibinfo{author}{Reeve, R.~M.} \&
  \bibinfo{author}{Kl{\"a}ui, M.}
\newblock \bibinfo{journal}{\bibinfo{title}{Development of a scanning electron
  microscopy with polarization analysis system for magnetic imaging with ns
  time resolution and phase-sensitive detection}}.
\newblock {\emph{\JournalTitle{Review of Scientific Instrum.}}}
  \textbf{\bibinfo{volume}{89}}, \bibinfo{pages}{083703}
  (\bibinfo{year}{2018}).

\bibitem{de2017modelling}
\bibinfo{author}{De~Clercq, J.}, \bibinfo{author}{Leliaert, J.} \&
  \bibinfo{author}{Van~Waeyenberge, B.}
\newblock \bibinfo{journal}{\bibinfo{title}{Modelling compensated
  antiferromagnetic interfaces with mumax3}}.
\newblock {\emph{\JournalTitle{Journal of Physics D: Applied Physics}}}
  \textbf{\bibinfo{volume}{50}}, \bibinfo{pages}{425002}
  (\bibinfo{year}{2017}).

\end{thebibliography}

\end{document}